# Ugo Fano and shape resonances

Antonio Bianconi

*Dipartimento di Fisica, Universit di Roma La Sapienza,
P.le Aldo Moro 2, 00185 Roma, Italy*

**Abstract.** Ugo Fano has been a leader in theoretical Physics in the XX century giving key contributions to our understanding of quantum phenomena. He passed away on 13 February 2001 after 67 years of research activity. I will focus on his prediction of the quantum interference effects to understand the photoabsorption cross section giving the Fano lineshapes . The Fano results led to the theoretical understanding of shape resonances (called also Feshbach resonances ) that should be better called Fano resonances . Finally I will show that today this Fano quantum interference effect is behind several new physical phenomena in different fields.

Ugo Fano was born in Turin, Italy, on 28 July 1912. His father Gino Fano (1871-1952) was professor of mathematics at Turin, Italy, specializing in differential geometry. He has spent his childhood mostly in Verona at "villa Fano" where he developed a love for mountains hiking and rock climbing. He studied Mathematics at University of Turin, but after the "Laurea" he turned his interests on physics having Enrico Persico (a Fermi's friend coming from Rome) as professor of theoretical physics and following the discussions with his cousin Giulio Racah (1909-1965), a theoretical physicist known for the powerful theory of angular momentum. He moved on 1934 to Rome to work with Enrico Fermi where he worked in the period 1934-1936. In these years the Fermi's group had shifted from atomic physics to the new experimental nuclear physics. They were doing systematic researches on the absorption and scattering properties of slow neutrons and discovered the artificial radioactivity induced by slow neutrons [1-3]. The fact that the neutron capture cross-section is high for small neutron velocities and the variation in different elements was interpreted as due to the capture of a neutron by a nucleus at a scattering resonance called "risonanza di forma" ("shape resonance").

Fano addressed his interests on the application of quantum theory for interpreting strange looking shapes of spectral absorption lines in the contiuum. Fano investigated the stationary states with configuration mixing under conditions of autoionization and he pointed out the basic physics of the quantum interference phenomenon between a discrete level and a continuum [4]. This theoretical result in atomic physics is related with the resonant scattering of a slow neutron in a nucleus, the "shape resonances", found by Fermi. In fact it deals with processes inverse to those considered in the Fano's theory of autoionization. This is a scattering process, in which the system is formed by combining an incident particle n (the neutron) with the "rest" (the nucleus) and then the system of N+1 particles breaks up releasing alternatively either the same particle or another particle and … . In this process the interference of resonance and potential scattering amplitudes gives a large reaction cross section. The theory of the nuclear scattering cross-section for neutron capture shows the same characteristic asymmetric "Fano lineshape" as it was shown by Breit and Wigner in 1936 [5].

In 1936-1937 Fano worked in Leipzig with Werner Heisenberg and he visited Arnold Sommerfeld, Niels Bohr, Edward Teller and George Gamow. From Germany Fano moved to Paris with the Joliot - Curie group. Fano returned to the University of Rome as a lecturer in 1938.

In these years Fano started to address his interests to radiation biology and genetics that will become central topics for his later research. It is noteworthy that, after a seminar in Rome by P. Jordan on x-ray effects on genetic material, Fermi had suggested to Fano that the biological action of radiation would be an important and suitable topic for study. He was in close contact with his school friend Salvatore Luria who moved from Turin to Rome where from 1935-1940 was in charge of Medical Physics and Radiology under the direction of Enrico Fermi and Edorado Amaldi. In 1939 Fano married Camilla ("Lilla") Lattes, who collaborated with him in science and worked as a teacher for many years. In the same year the couple immigrated to the United States in 1939 to escape the racial laws.

He was at the University of Michigan summer school at Ann Arbor in 1939 when Werner Heisenberg and Edoardo Amaldi visited Fermi in the States. Once he told me that during a party in August he became aware that for everybody there it was clear that Fermi and Heisenberg would become the leaders of the USA and German projects for the nuclear bomb. At this point he decided not to work in nuclear physics being interested on other science issues. He focused on the interaction of Radiation with matter and in particular on effects of radiation on living organisms (genetic resistance to radiation effects). In 1940-1944 worked with the help of his wife in what was later to be called radiation biology at the Department of Genetics of the Carnegie Institution at Cold Spring Harbor. Fano's papers in this period concerned chromosomal rearrangement mutations, lethal effects, and genetic effects of X-rays and neutrons on Drosophila melanogaster, as well as theoretical analysis of genetic data. His work also included the discovery of bacteriophage-resistant mutants in Escherichia coli, following up earlier studies by Salvador E. Luria who moved to USA in 1939 and visited the Carnegie Institution.

After the war in the years1946-1966, he joined the staff of National Bureau of Standards (NBS), initially working in the radiological physics group led by Lauriston S. Taylor and after in the basic physics of atoms, molecules, and condensed matter elucidating fundamental physical processes.

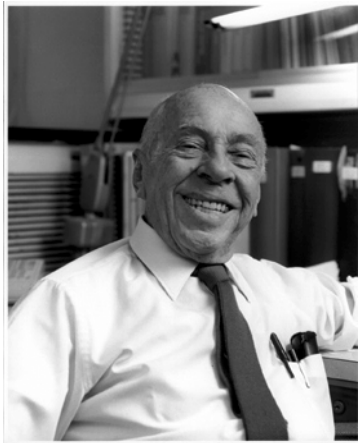

**FIGURE 1.** Ugo Fano

In the fifties the scientific interest for shape resonances was coming up again, in fact the availability of monochromatic neutron beams at nuclear reactors allowed the measure of the neutron cross section on selected nuclei as a function of neutron energy. The trapping of the slow neutrons for long time inside the nuclei has been clearly shown to be at the origin of the shape resonances for neutron capture. The theory of "shape resonances" in the frame of potential scattering was developed by J. M. Blatt and V. F. Weisskopf [6] and later by Herman Feshbach in 1958 [7] for a general nuclear reaction theory, based on the projection of the nuclear state into direct and compound channels, following the method introduced by Fano [4].

The shape resonance occurs when a quantum particle with energy E and wave vector $k=2\pi/\lambda$ is trapped within a potential well with finite barrier of a given size R given by the radius of the nucleus, with the generic condition for the shape resonance: $R = n\lambda/2$ where n is an integer. The name "shape resonance" indicates the fact that the shape of the potential barrier determines the energy of the resonance, therefore this resonance has been used in 1949-1954 by Edoardo Amaldi and others to measure the size of the nuclei of several elements with the precision of $-10^{-13}$ cm [8].

In these years the interest of Fano returns to this phenomenon following the extensive investigation of line profiles high energy levels of excitation in the far-UV absorption spectra in atomic and molecular spectroscopy undertaken by means of far-ultraviolet light of electron bombardment and also of energy transfer in molecular collisions. The Fano prediction and interpretation of the experiments on the excitation of quasi bound states buried in continua whose spectra was his major outcome in these years [9] and the review paper on the *The theory of atomic photoionization"* remains as a relevant milestone in the physics of XX century [10]. He has shown that lineshape of the absorption lines (Fig. 2) in the ionization continuum of atomic (and molecular) spectra are represented by the formula

$$\sigma(\varepsilon) = \sigma_a \frac{(q+\varepsilon)^2}{1+\varepsilon^2} + \sigma_b \tag{1}$$

where $\varepsilon = \dfrac{E-E_r}{\Gamma/2}$ indicates the deviation of the incident photon energy E from the idealized resonance energy $E_r$ which pertains to a discrete auto-ionizing level of the atom. This deviation is expressed in a scale whose unit is the half-width $\Gamma/2$ of the line ($h/\Gamma$ is the mean life of the discrete level with respect to autoionization). $\sigma(\epsilon)$ represents the absorption cross section for photons of energy E whereas $\sigma_a$ and $\sigma_b$ are two portions of the cross section corresponding to transitions to states of the continuum that do and do not interact with the discrete auto-ionizing state respectively. Finally q is a numerical index which characterizes the line profile.

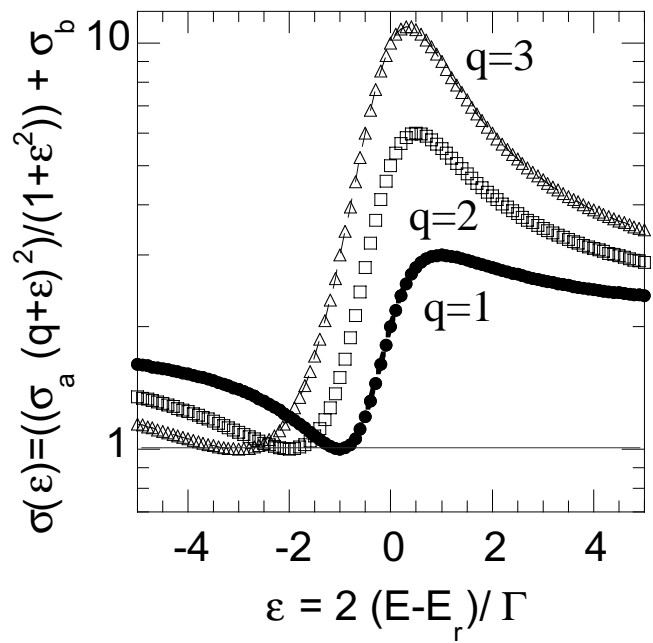

**FIGURE 2.** The Fano lineshape of the absorption cross near resonance energy $E_r$ of a discrete state buried in the continuum for various values of the q parameter.

Thus, he encouraged Robert P. Madden and co-workers at NBS to use synchrotron radiation for spectroscopic studies. He convinced his old friend Edoardo Amaldi and Mario Ageno to push the Italian scientific community to use the Frascati synchrotron as a synchrotron radiation source for high energy spectroscopy of atoms and solids.

In 1966 he joined the Physics faculty of Chicago where he continued his research on the interaction of radiation with matter. I was still in my twenties when I had the possibility to meet him regularly in the years 1972-1975. He used to come to Rome in June-July, spending his time in Frascati Laboratories with the small group of synchrotron radiation researchers: Adalberto Balzarotti, Emilio Burattini, Mario Piacentini and myself. I remind very nice days discussing the x-ray absorption spectra measured at the new Frascati soft x-ray synchrotron radiation beam line. He informed me on the resonances observed in the scattering of electrons on nitrogen molecule that were described by Dehmer and Dill with the same formalism of his "shape resonances". These discussions led me to the interpretation of the x-ray absorption near edge structure (XANES) in complex solids and metalloproteins in term of "shape resonances" of the excited photoelectron within a finite cluster of atoms surrounding the absorbing atom [11, 12].

The Fano quantum interference effects have been observed in modern experiments of photo-fragmentation [13]. Here the Fano quantum interference appears when a quantum state is formed at the resonance energy $E_r$ above the fragmentation threshold of the system. The variation of the cross section in the neighborhood of the energy of fragmentation resonance follows the Fano lineshape. In the field of photo-fragmentation these resonances are usually classified into "Feshbach resonances" if several (n>1) electronic transitions are required to emit one electron or "shape resonances" if only one electronic transition is required to emit one electron. However this classification makes sense only if the lifetimes of the resonances are not larger than the typical vibrational time of the system, otherwise the distinction is not clear. The quantum mechanical phase shift and consequent interference effects encountered on passing through a shape resonance energy have been studied by many authors and they can be seen as a movie on the web [14].

In 1963 it was pointed out by Thompson and Blatt [15] that the basic physics of the quantum interference between a discrete state and a continuum introduced by Fano predicts the amplification of the superconducting critical temperature in a superconducting film of thickness R if R=$\lambda_F$/2 x integer, where $\lambda_F$ is the wavelength of the electrons at the Fermi level. This prediction did not work since in a single film phase fluctuations suppress the condensate phase coherence. In 1993 we have shown that the shape resonance amplification works in a superlattice of superconducting wires [16] where the enhancement is obtained by tuning the Fermi level at a shape resonance of the superlattice of period L=$\lambda_F$/2 x integer. The quantum interference effects between the pairs in a narrow band and in a wide band give two superconducting gaps in two different bands in momentum space and in real space, that has been recently confirmed in the new $MgB_2$ superconductor made by a superlattice of boron layers [17].

In 1995 the Bose Einstein condensation (BEC) of atoms trapped magnetically in vacuum and cooled to a few billions of a degree above absolute zero has been achieved [18]. This ultracold atomic gas is a dilute system in which the interparticle interactions are weak and easy to be treated theoretically. It has been found that using a magnetic field it is possible to control the trapping of the free atoms into molecular diatomic molecules for a short time that are known as Feshbach resonances (19) but they will be better called Fano resonances and it is possible to control the fundamental interactions (20). Moreover the inhomogeneous trapping potential leads to spatial separation of high- and low-energy atoms, giving rise to a Fermi surface that is manifest in the real as well as in the momentum space. Recently the onset of Fermi degeneracy in a ultracold gas of fermioms has been reported [21]. Here the pairing interaction between fermions could be controlled by tuning the system to a Feshbach resonance and the condensation of pairs (as in He $_3$) could become possible, similar to Cooper pair formation in superconductivity. This process is similar to the $T_c$ amplification by shape resonance in high $T_c$ superconductors [16].

The Fano lineshapes have been seen also in the zero bias conductance as a function of the gate voltage in a single electron transistor [22]. Here an artificial atom is created by making a layer of GaAs on top of which is a layer of AlGaAs doped with Si. The electrons from the dopants fall into the GaAs, and the resulting positive charge on the Si atoms creates a potential that holds the electrons at the GaAs/AlGaAs interface, creating a two dimensional electron gas. The quantization of energy and charge makes the confined droplet of electrons closely analogous to an atom. The resonance component in the Fano interference appears to come from the single electron charging of the artificial atom interacting with a continuous component.

In conclusion I have shortly focused only on a particular aspect [4, 9, 10] of the wide scientific activity of Ugo Fano that had a relevant influence on my personal scientific activity and today it is stimulating new physics.